\documentstyle[float,preprint,tighten,aps,pre,amsfonts,amssymb]{revtex}

\input{epsf}

\newcommand{\mean}[1]{\left\langle{#1}\right\rangle}

\newcommand{\SNR}{\text{SNR}}  

\renewcommand{\vec}[1]{{\bf #1}}
\newcommand{\mat}[1]{{\bf #1}}

\newcommand{\Dpsi}{\Delta\psi}

\newcommand{\dgr}[1]{#1^{\circ}}

\newcommand{\eenum}[2]{#1 \cdot 10^{#2}}

\newcommand{\Ttrans}[2]{{\cal T}\left(#1\,|\,#2\right)}
\newcommand{\phasd}[2]{\chi^{(#1)}\left(#2\right)}               
\newcommand{\PD}[1]{\bbox{\chi}^{(#1)}}

\renewcommand{\Re}{\text{Re}}

\begin{document}
\draft

\title{Markov analysis of stochastic resonance in a periodically
       driven integrate-fire neuron}

\author{Hans E.\ Plesser%
        \footnote{Electronic address: \texttt{plesser@chaos.gwdg.de}}
        and Theo Geisel}       
\address{Max-Planck-Institut f\"ur Str\"omungsforschung
         and Fakult\"at f\"ur Physik, Universit\"at G\"ottingen,
         Bunsenstra\ss{}e~10, 37073~G\"ottingen, Germany}

\date{\today}
\maketitle

\begin{abstract}
  We model the dynamics of the leaky integrate-fire neuron under
  periodic stimulation as a Markov process with respect to the
  stimulus phase.  This avoids the unrealistic assumption of a
  stimulus reset after each spike made in earlier work and thus solves
  the long-standing reset problem. The neuron
  exhibits stochastic resonance, both with respect to input noise
  intensity and stimulus frequency.  The latter resonance arises by
  matching the stimulus frequency to the refractory time of the
  neuron.  The Markov approach can be generalized to other
  periodically driven stochastic processes containing a reset
  mechanism.
\end{abstract}




\section{Introduction}\label{sec:intro}

Periodically modulated stochastic processes have been studied
intensely over the last two decades under the paradigm of stochastic
resonance: the transduction of signals is optimal in the presence of a
particular amount of noise.  First suggested to explain the
periodicity of ice-ages~\cite{Benz:1981(L453)}, stochastic resonance
has since been demonstrated in a wide range of experiments and the
underlying mechanisms are well understood.  A recent review of the
field is given in~\cite{Gamm:1998(223)}.

The concept of stochastic resonance has met with particular attention
in the
neurosciences~\cite{Long:1993(309),Wies:1995(33),Levi:1996(165),%
Coll:1996(642),Cord:1996(769)}.  The brain achieves an enormous signal
processing performance in the presence of noise from a wide range of
sources, ranging from stochastic membrane channel openings on a
molecular level, via highly irregular firing patterns of individual
neurons to distracting stimuli in perception.  The improvement of
signal transduction on all of these levels has now been demonstrated
experimentally~\cite{Bezr:1997(2456),Doug:1993(337),%
Stem:1995(1877)}.  Recently, the first direct evidence for the
behavioral relevance of stochastic resonance has been
reported~\cite{Moss:1998(CNS)}, underlining the importance of
stochastic resonance in neurobiology.

In short, neurons are threshold devices that receive an input $I(t)$
which charges the membrane of the neuron like a leaky capacitor.  When
the potential $v(t)$ across the membrane reaches a threshold $\Theta$,
a \emph{spike} is fired: the membrane potential makes a brief but
strong excursion (duration $\approx 2$ms, amplitude $\approx 100$mV).
This spike is transmitted as output to other neurons.  After the
spike, the membrane potential is reset to a resting value $v_0$, some
$30$mV below the threshold~\cite{Nich:From(1992)}.  As the shape of
the spikes is stereotypical, information is only conveyed by the spike
times.

This has led to the leaky integrate-fire model of neuronal
dynamics~\cite{Tuck:Stoc}.  In between two spikes, the membrane
potential is governed by 
\begin{equation}
  \tau_m \dot{v}(t) = -v(t) + I(t) + \zeta(t) \:.
  \label{eq:if_model}
\end{equation}
Here, $\tau_m$ is the time-constant of the membrane, which represents
the internal time-scale of the neuron and $\zeta(t)$ is an as yet
undefined noise process, comprising, e.g., stochastic membrane
potential fluctuations and irregular input to the neuron from sources
uncorrelated to $I(t)$.  As the potential reaches the threshold, a
spike is recorded and the potential is reset to $v(t)=v_0$
instantaneously.

For Gaussian white noise $\zeta(t)$ the evolution of the membrane
potential $v(t)$ from reset potential to threshold is equivalent to an
Ornstein--Uhlenbeck process with drift $I(t)$ and an absorbing
boundary at $v=\Theta$.  The output of the neuron is modeled as a
sequence of delta pulses $f(t) = \sum_{k} \delta(t-t_k)$ at the times
of threshold crossings $\{t_k\} = \{t|v(t)=\Theta\}$ (\emph{spike
train}).  The spike train is a stochastic point process, specified
entirely by the spike times $\{t_k\}$.

This biologically most interesting stochastic process has so far
escaped a rigorous analysis, in spite of several partially successful
attempts~\cite{Buls:1996(3958),Ples:1997(228)}.  For a list of open
issues see Sec.~V.C.4 of the review by Gammaitoni et
al.~\cite{Gamm:1998(223)}.  This is in marked contrast to the
treatment of mathematically more accessible, but biologically less
plausible models, such as bistable dynamic
systems~\cite{McNa:1989(4854),%
Jung:1993(175),Zhou:1990(3161),Buls:1991(531)} and threshold devices
\emph{without
reset}~\cite{Ging:1995(191),Jung:1994(2513),Wies:1994(2125)}, in which
stochastic resonance has been well established.

The essential difficulty arises from the reset after each spike: there
is no well-defined membrane potential distribution for asymptotic
times, as used in the case of reset-free threshold detectors.  Instead
we have to analyze each inter-spike-interval separately and then put
these pieces together to obtain the spike train as a whole.  To
facilitate this, past work has assumed that the durations of all
inter-spike-intervals ($\tau_k = t_k-t_{k-1}$) were identically and
independently distributed (i.i.d.), i.e.\ that the spike train is a
stationary renewal process~\cite{Cox:Stat}.  But in the presence of
time-dependent input $I(t)$, this would require \emph{identical input}
within each inter-spike-interval (ISI).  This is the much criticized
reset assumption: If the model neuron were to describe a neuron in the
auditory nerve while you are listening to a music tape, the reset
assumption requires that upon the firing of each spike the tape should
be rewound to exactly the position it had at the time of the last
spike!

In this work we show how to analyze the response of the leaky
integrate-fire neuron to periodic stimuli without undue
assumptions.  The distribution of the length of individual inter-spike
intervals is computed numerically~\cite{Ples:1997(228)}, and spike
trains are then assembled as Markov chains from these intervals.  We
obtain probability distributions for the length of inter-spike
intervals and the stimulus phases at which spikes occur.  These
distributions should be directly comparable to experiments employing
sustained stimulation with periodic signals.  The signal processing
performance of the neuron is judged by the signal-to-noise ratio (SNR)
of the output spike train.  The SNR is maximal at an optimal noise
amplitude for fixed stimulus frequency and at a resonance frequency
for fixed noise amplitude.  The latter resonance is a consequence of a
time-scale matching between stimulus and membrane time-constant.  All
computations are verified by simulations.

In Sec.~\ref{sec:markov}, we show how to exploit the Markov property
of the integrate-fire neuron to determine its response to sinusoidal
input $I(t)$.  The performance of the model neuron as a signal
processing device is investigated in Sec.~\ref{sec:sr}.  The results
are discussed in Sec.~\ref{sec:disc}.


\section{Markov analysis}\label{sec:markov}

For an input current consisting of a constant offset and a sinusoidal
component, and Gaussian white noise the Langevin
equation~(\ref{eq:if_model}) reads 
\begin{equation}
  \dot{v}(t) = -v(t)  + \mu + q \cos(\Omega t +\phi_0) + \sqrt{D} \xi(t) \;,
  \label{eq:if_scaled}
\end{equation}
where time and potential have been scaled to their respective natural
units $\tau_m$ and $\Theta$; the reset potential is set to $v_0 = 0$.
The input is characterized by the DC offset $\mu$, stimulus amplitude
$q$, frequency $\Omega$ and initial phase $\phi_0$.  The noise has
amplitude $\sqrt{D}$ and autocorrelation
$\mean{\xi(t)\xi(t')}=\delta(t-t')$.  In the remainder of this
article, we will investigate this model.  For a derivation of the type
of input current used here from more elementary models,
see~\cite{Lans:1997(2040)}.

In the absence of noise ($D=0$), spikes will only be generated if
\[
  v_{\infty}=\lim_{t\to\infty} v(t) 
     = \mu +\frac{q}{\sqrt{1+\Omega^2}} > 1 \;.
\]
Therefore, we classify stimuli as sub-threshold if $v_{\infty} \le 1$
and as supra-threshold otherwise.  In this work, we will focus on the
biologically more interesting sub-threshold
regime~\cite{Kemp:1998(1987)}. The methods presented here are
applicable independent of the choice of stimulus parameters.  We only
require the presence of noise, i.e.\ $D>0$.

Suppose that an initial spike has occured at time $t_0 = 0$,
corresponding to stimulus phase $\phi_0$. The next spike follows at
time $t_1 = \inf\{t>t_0|v(t)\ge 1\}$ and stimulus phase $\phi_1 =
(\Omega t_1+\phi_0) \bmod 2\pi$, whence $\cos(\Omega (t-t_1) +\phi_1)
= \cos(\Omega t +\phi_0)$ for $t>t_1$. This suggests to re-write
Eq.~(\ref{eq:if_scaled}) in terms of the time $t'$ that has passed
since the most recent spike at phase $\phi$.  Thus, given this phase,
the potential evolves from $v(t'=0|\phi)=v_0=0$ until the next spike
according to
\begin{equation}
   \dot{v}(t'|\phi) = -v(t'|\phi) + \mu + q \cos(\Omega t'+\phi) 
                     + \sqrt{D}\xi(t') \;.
   \label{eq:langevin}
\end{equation}
The next spike is fired after an interval $\tau$, as soon as the
threshold condition is met
\begin{equation}
   \tau = \inf\{t'>0|v(t'|\phi)\ge1\} \;.
   \label{eq:threshold}
\end{equation}
The inter-spike intervals are connected by the iteration equations 
\begin{equation}
   \phi_k = (\Omega \tau_k+ \phi_{k-1}) \bmod 2\pi  \;,\quad
    t_{k} = t_{k-1} + \tau_k \;,
   \label{eq:iterate}
\end{equation}
leading to the output spike train 
\begin{equation}
  f(t) = \sum_{j=0}^{\infty} \delta(t-t_j) 
       = \sum_{j=0}^{\infty} \delta\bigg(t-\sum_{k=1}^{j}\tau_k\bigg) \;.
  \label{eq:train}
\end{equation}

The reset of the membrane potential to $v_0=0$ after each spike
completely erases the memory of the neuron.  The subsequent behavior
of the neuron therefore depends on its past only through the absolute
time of the spike $t_k$, i.e.\ the spike train is a Markov process.

We have thus split the task of solving the dynamics of the
integrate-fire neuron into two parts.  We will first solve the
first-passage-time problem posed by Eqs.~(\ref{eq:langevin})
and~(\ref{eq:threshold}) for a given phase $\phi$ of the last spike,
before assembling the spike train from the inter-spike intervals
according to Eqs.~(\ref{eq:iterate}) and~(\ref{eq:train}).

\subsection{Conditional ISI distribution}\label{ssec:cISI}

The first-passage-time problem for the membrane potential posed by
Eqs.~(\ref{eq:langevin}, \ref{eq:threshold}) yields the distribution
$\rho(\tau|\phi)$ of the inter-spike-interval lengths $\tau$ for a
given stimulus phase $\phi$ at the beginning of the interval
(conditional ISI distribution).  To the best of our knowledge, no
analytic solution is known for this seemingly simple
first-passage time problem of the Ornstein--Uhlenbeck process.  The
approximations suggested in~\cite{Buls:1996(3958)} are valid in a
restricted parameter range only---low stimulus frequencies in
particular---and appear to yield qualitative rather than quantitative
agreement with simulations.  

We employ here a numerical method to compute the inter-spike-interval
distributions.  The method is discussed in detail
in~\cite{Ples:1997(228)}, and we only sketch it here.  In the
absence of an absorbing threshold the probability 
${\cal P}(w,t'|u,s';\phi)$ that the membrane potential is
$w$ at time $t'$ if it was $u$ at time $s'<t'$ is a Gaussian
distribution.  The mean is given by the solution at time $t'$ of
Eq.~(\ref{eq:langevin}) for the noise-free case ($D=0$) with initial
condition $v(s'|\phi)=u$, while the variance is 
$\sigma^2(t')=\frac{D}{2}(1-e^{-2(t'-s')})$.  Then, the
inter-spike-interval distribution is given by the integral
equation~\cite{Kamp:Stoc(1992)}
\begin{equation}
  {\cal P}(1, t' | 0, 0) = 
     \int_0^{t'} {\cal P}(1, t' | 1, \tau) \rho(\tau|\phi) d\tau \;.
  \label{eq:volterra}
\end{equation}
This equation is solved for $\rho$ using standard
techniques~\cite{Linz:Anal}. Source code is available on request.

\begin{figure}             
  \centerline{\epsfbox{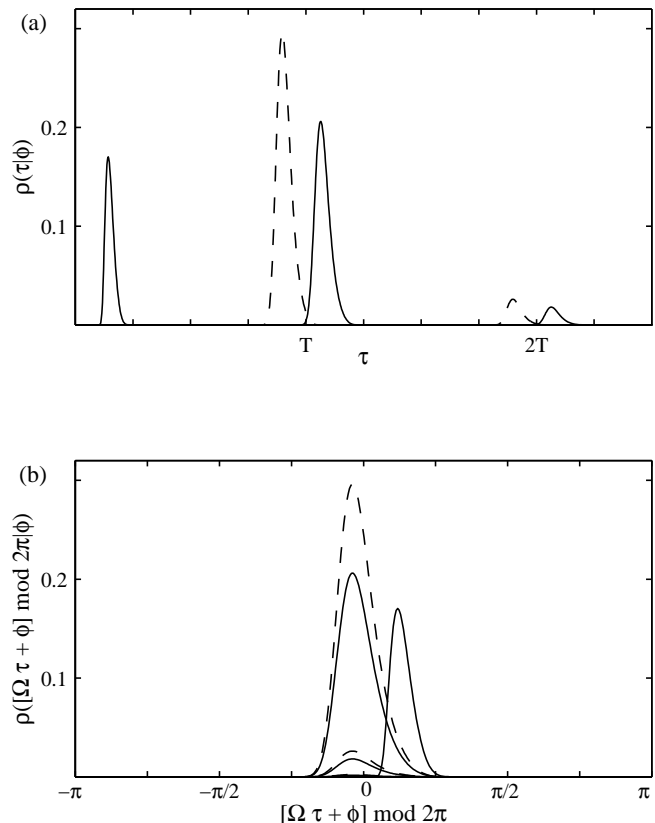}}
  \caption{%
    (a)~Conditional ISI distributions for $\phi=-\pi/6$
        (solid) and $\phi=\pi/6$ (dashed); other parameters $\mu=0.95$,
        $q=0.048$, $\Omega=0.05\pi$, $D=\eenum{6}{-5}$.  
        $T=40$ is the stimulus period.  
        The refractory mode at small $\tau$  is present only
        for $\phi=-\pi/6$.  The small modes around
        $2T$ correspond to the probability of skipping a period.
    (b)~The same distributions as in (a), but now plotted vs.\ phase,
        $\psi = [\Omega\tau+\phi] \bmod 2\pi$, shifted to $[-\pi, \pi]$.
        The modes at $\approx -\pi/25$ coincide for the first and
        second stimulus period, while the refractory mode is
        clearly set apart.}
\label{fig:cISI}
\end{figure}

As shown in Fig.~\ref{fig:cISI}, the conditional inter-spike-interval
distributions $\rho(\tau|\phi)$ may depend strongly on $\phi$.  First,
they contain a series of exponentially decaying peaks that are
separated by the stimulus period $T=2\pi/\Omega$.  These peaks
represent spikes that are well phase-locked to the stimulus and we
will refer to them as \emph{periodic peaks}.  An additional peak
appears at short intervals $\tau$ for certain phases $\phi$.  This
peak reflects the rise time of the membrane potential towards
threshold.  Its location is not related to the stimulus period $T$,
but reflects the intrinsic time-scale of the neuron and defines its
refractory time, i.e.\ the minimum interval between two spikes.  Thus,
we will refer to this peak as the \emph{refractory peak}.  It
corresponds to two or more spikes fired in rapid succession within a
single stimulus period (a \emph{burst}).  There is thus a qualitative
dependence of the distributions $\rho$ on the phase $\phi$ that can
lead to interesting consequences for the firing behavior of the
neuron.

The relation between periodic and refractory peaks depends
on the stimulus parameters, particularly on the frequency and the noise
amplitude.  We will discuss this relationship in Sec.~\ref{ssec:stat}.

\subsection{Markov process in phase}\label{ssec:chain}

Let us now turn to the problem of assembling spike trains from
inter-spike-intervals according to Eqs.~(\ref{eq:iterate},
\ref{eq:train}).  The length of an interval following a spike at time
$t$ and stimulus phase $\phi=[\Omega t + \phi_0] \bmod 2\pi$ is
distributed according to $\rho(\tau|\phi)$.  Therefore, the
probability that the next spike will occur at phase $\psi$ is given by
\begin{equation}
  \Ttrans{\psi}{\phi} = \int_0^{\infty} 
   \rho(\tau|\phi) \delta(\psi - [\Omega\tau+\phi] \bmod 2\pi) 
   \frac{d\tau}{\Omega} \;.
  \label{eq:Tdef}
\end{equation}
We will call $\Ttrans{\psi}{\phi}$ the transition probability of the
spike phase.  We will now consider the Markov process of the spike
phases $\phi_k$ instead of the Markov process made up of the spike
times $t_k$.

If we define the spike phase distribution $\phasd{k}{\phi}$ as the
probability (across an ensemble of neurons or repetitions of an
experiment) that the $k^{\text{th}}$ spike in a train will be fired at
stimulus phase $\phi$, then this probability will evolve according to
\begin{equation}
  \phasd{k+1}{\psi} = \int_0^{2\pi}
              \Ttrans{\psi}{\phi} \phasd{k}{\phi} d\phi \;.
  \label{eq:chi_evol}
\end{equation}
As the neuron fires repetitively while driven by a stationary periodic
stimulus, the spike train emitted by the neuron will approach a stationary
Markov process with phase distribution
\begin{equation}
   \phasd{s}{\psi} 
   = \lim_{k\rightarrow\infty} \phasd{k}{\psi}  
    = \int_0^{2\pi} \Ttrans{\psi}{\phi} \phasd{s}{\psi} d\phi \;.
  \label{eq:chi_eig}
\end{equation}
The stationary phase distribution $\phasd{s}{\psi}$ is the
eigenfunction to eigenvalue $1$ of the kernel $\Ttrans{\psi}{\phi}$,
and is guaranteed to exist because this kernel is a conditional
probability distribution~\cite{Bake:Nume}.  Any initial phase
distribution will converge to the unique stationary solution provided
that $\Ttrans{\psi}{\phi}>0$ everywhere~\cite{Mise:Math}.  That the
latter condition holds in the presence of noise can be seen as
follows.  For sub-threshold stimuli, noise may drive the potential
across the firing threshold at any time $\tau>0$ in principle,
yielding a possibly tiny, but non-zero probability of spikes at any
phase.  The same argument holds true for supra-threshold stimuli,
where noise may keep potential below threshold up to any time.  In the
absence of noise, neither convergence nor uniqueness are assured.

To facilitate numerical treatment, we discretize the phase.  Since the
conditional inter-spike-interval distributions $\rho(\tau|\phi)$ are
smooth in both time and phase due to the presence of noise in the
input, this discretization will introduce only minor numerical errors.
It is largely equivalent to applying numerical methods to solve the
kernel eigenvalue problem~\cite{Bake:Nume}.  Using $L$ bins of width
$\Dpsi$ ($\Dpsi=2\pi/L$) we obtain the spike phase distribution vector
\begin{equation}
   \bbox{\chi} = 
          \left(\chi_0, \chi_1, \ldots, \chi_{L-1}\right)^{\text{tr}}
   \;,\quad
   \chi_j  = \int_{j\Dpsi}^{(j+1)\Dpsi} \chi(\psi) d\psi \;,
   \label{eq:chid}
\end{equation}
and the phase transition matrix $\mat{T}$ with elements
\begin{equation}
   \mat{T}_{jk} = \int_{j\Dpsi}^{(j+1)\Dpsi} 
                            \Ttrans{\psi}{k\Dpsi} d\psi
   \;,\quad j, k = 0, \ldots, L-1 .
   \label{eq:Td}
\end{equation}
The evolution equation (\ref{eq:chi_evol}) simplifies from convolution to
matrix-vector multiplication  
\begin{equation}
   \PD{k+1} = \mat{T} \cdot \PD{k} \;,
   \label{eq:chid_evol}
\end{equation}
and the stationary distribution $\PD{s}$ is the eigenvector to
eigenvalue $1$ of the matrix $\mat{T}$.  We have thus reduced the
Markov process to a Markov chain.  

In practice, we obtain the transition matrix $\mat{T}$ by numerically
evaluating equations~(\ref{eq:Tdef}) and~(\ref{eq:Td}), with
$\rho(\tau|\phi)$ from Eq.~(\ref{eq:volterra}).  The stationary
distribution is then found using standard eigenvector routines.  For
all data shown here, we used the discretization $L=72$, $\Dpsi =
\pi/36 = \dgr{5}$. In figures of transition matrices and phase
distributions the axis will run from $-\pi$ to $\pi$ as this renders
structures more clearly.

\begin{figure}                
  \centerline{\epsfbox{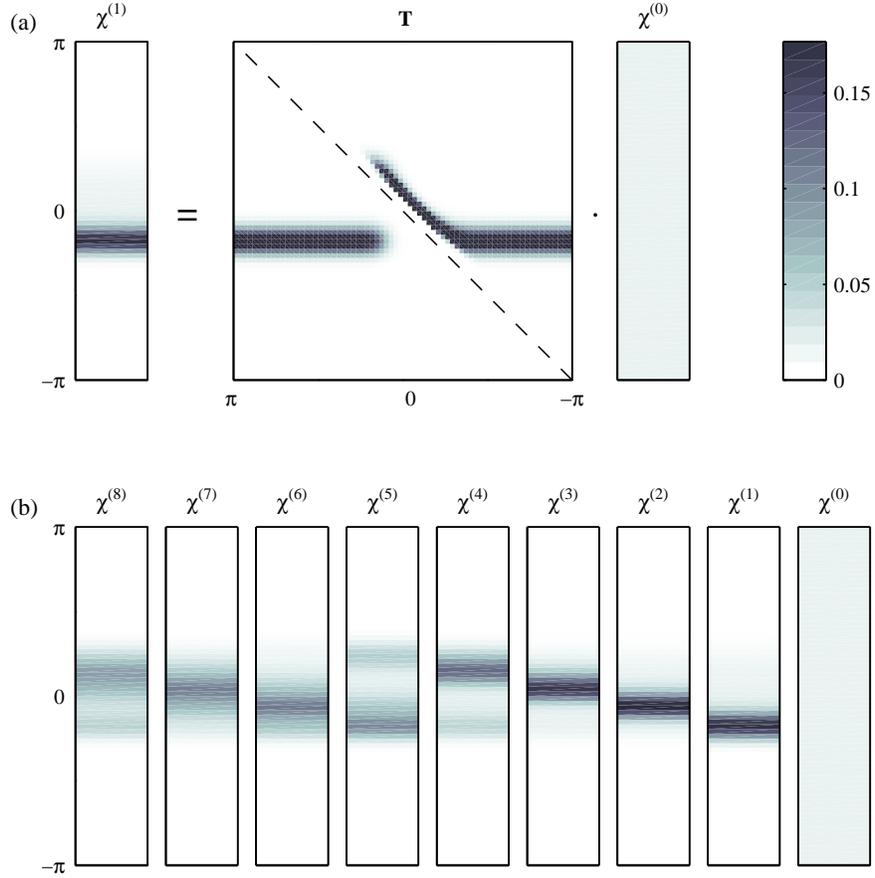}}
  \caption{%
      (a)~Graphic representation of the Markov chain iteration 
          given by Eq.~(\ref{eq:chid_evol}).  The dashed
          line is the matrix diagonal.  Probability is given by
          grayscale as indicated by the colorbar.  
      (b)~Evolution of an initially uniform phase distribution under 
          subsequent 
          multiplications with \mat{T}, from right to left.  
       	  See text for details.
          Stimulus parameters:
          $\mu=0.95$, $q=0.05$, $\Omega=0.02\pi$, $D=\eenum{1.3}{-4}$.}
\label{fig:mc_sample}
\end{figure}

An example for the phase evolution of an initially uniform
distribution towards the stationary state under the influence of a
transition matrix $\mat{T}$ is given in Fig.~\ref{fig:mc_sample}.  To
``read'' the transition matrix, note that the matrix columns
correspond to the phase $\phi_k$ of the spike preceding the interval,
the rows to the phase $\phi_{k+1}$ of the spike terminating it.  The
phase axes run from $-\pi$ to $\pi$ from bottom to top in phase
distribution vectors $\bbox{\chi}$ and the rows of the transition
matrix $\mat{T}$, and \emph{from right to left} across the columns of
$\mat{T}$.  Thus, the horizontal bar in the transition matrix shown in
Fig.~\ref{fig:mc_sample} indicates that for most values of $\phi_k$
the next spike will occur around $\phi_{k+1}\approx -\pi/6$.  This bar
corresponds to the periodic peaks of the ISI distributions.  For
$-\pi/4 \lesssim \phi_k \lesssim\pi/6$, the matrix is dominated by a
``finger'', running parallel to the matrix diagonal.  Within this
range of phases, a spike will be followed by another spike at a
slightly later phase, as shown in Fig.~\ref{fig:mc_sample}b.
Figuratively speaking, the neuron fires a burst of spikes, but there
is always a chance that two subsequent spikes will be one or more
stimulus periods apart, even though they are close in phase: in the
Markov chain description, all information about actual interval
lengths is lost.  The finger results from the refractory peak of the
ISI distributions.

Figure~\ref{fig:chis_low} shows the dependence of transition matrix and
stationary phase distribution on the noise amplitude for slow stimuli
($T\gtrsim 10$).  For low noise, the transition matrix is dominated by the
horizontal bar, which intersects with the matrix diagonal, indicating a
stochastic fixed point.  This results in a sharply peaked spike
phase distribution.  At intermediate noise, the finger is more
pronounced, while the bar barely touches the matrix diagonal, leading to a
stochastic limit cycle with two preferred phases: the neuron often fires
bursts of two successive spikes.  At high noise, the finger stretches all
along the matrix diagonal, while the horizontal bar has disappeared
altogether.  The neuron fires rapidly, but largely uncorrelated with the
stimulus and the phase distribution is virtually flat.

This means that for very low noise the spike train of the neuron is
nearly a stationary renewal process with inter-spike-intervals i.i.d.\
according to $\rho(\tau|\psi^{*})$.  Here $\psi^{*}$ is the location
of the maximum of the stationary phase distribution, which depends not
only on the stimulus parameters, but also on the noise amplitude.  For
high noise amplitudes, the response of the neuron is largely
independent of the stimulus, and may thus be described by a stationary
renewal process as well---the ISIs reduce to the refractory peak.  But
at intermediate noise levels---i.e.\ those essential to the
observation of stochastic resonance---the stationary phase
distribution may be multimodal.  Thus the correlations between the
phases of subsequent spikes have to be taken into account using the
Markov ansatz.  Multimodal phase distributions as discussed here are
not just hypothetical: they have been observed in sensory neurons of
goldfish upon stimulation with sinusoidal water
waves~\cite{Mogd:1998(priv)}.

For fast stimuli ($T\lesssim 10$), the stationary phase distribution
smears out much more along the phase axis, and does not show
multimodality, because the refractory time of the neuron becomes
comparable to the stimulus period and bursting is no longer possible,
see Fig.~\ref{fig:chis_high}.  At low to intermediate noise, the
distribution is too wide to be replaced by its mode as in the renewal
ansatz, but still sufficiently narrow to provide for a response that
is well phase-locked to the stimulus.  Therefore, the Markov approach
is essential for high frequency stimuli as well.

\subsection{Stationary ISI distribution}\label{ssec:stat}

Once the stationary phase distribution is known, the inter-spike-interval
distribution of the stationary firing process is obtained by averaging the
conditional ISI distributions over phase
\begin{equation}
  \rho(\tau) = \int_0^{2\pi} \rho(\tau|\psi) \phasd{s}{\psi} d\psi \;.
  \label{eq:sISI}
\end{equation}
The average interval length thus is
\begin{equation}
  \mean{\tau} = \int_0^{\infty} \tau \rho(\tau) d\tau \;.
  \label{eq:mfpt}
\end{equation}
$\rho(\tau)$ is the inter-spike-interval distribution that we expect
to find in experiments with tonic stimulation.  In contrast to a
stationary renewal process, this averaged ISI distribution does
\emph{not} contain a full description of the spike train.

\begin{figure}                    
  \centerline{\epsfbox{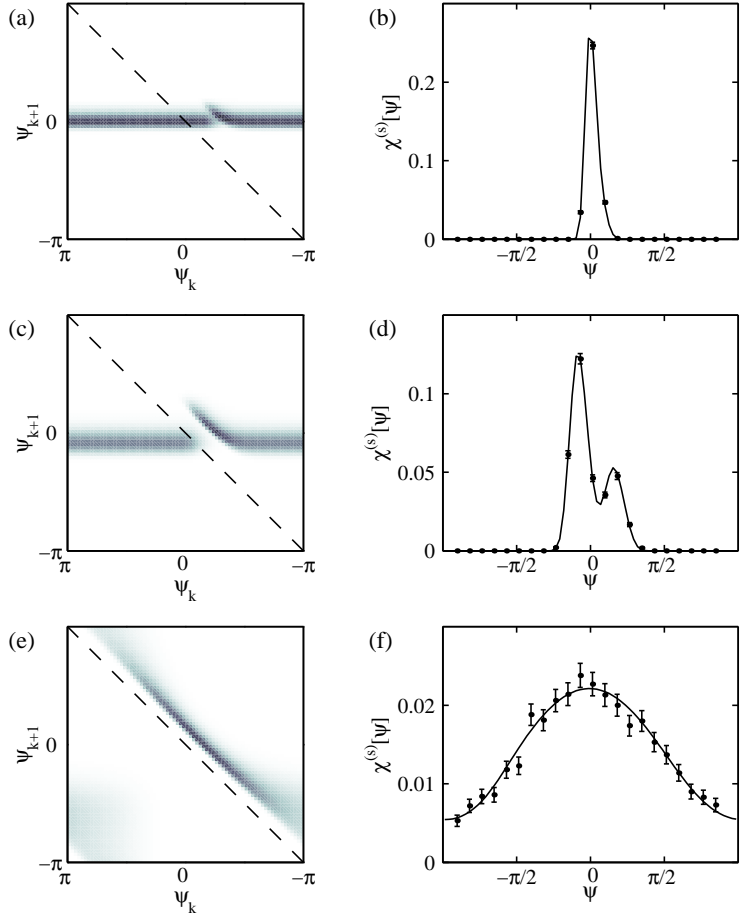}}
  \caption{%
  Phase transition matrices $\mat{T}$
  (a,~c,~e) and corresponding stationary phase distributions $\PD{s}$
  (b,~d,~f) for stimulus frequency $\Omega = 0.05\pi$ 
  at three different noise intensities $D = \eenum{6.2}{-6}$ (a,~b),
  $D = \eenum{7.0}{-5}$ (c,~d), and $D = \eenum{4.8}{-3}$ (e,~f);
  other parameters: $\mu = 0.95$, $q = 0.05$.  The grayscale is the
  same for all matrices, white indicating vanishing probability.
  Error bars in the phase distributions indicate standard error of
  mean from simulated trains of 20,000 spikes.  
  Observe the different scalings of the ordinate.}
\label{fig:chis_low}
\end{figure}

\begin{figure}                   
  \centerline{\epsfbox{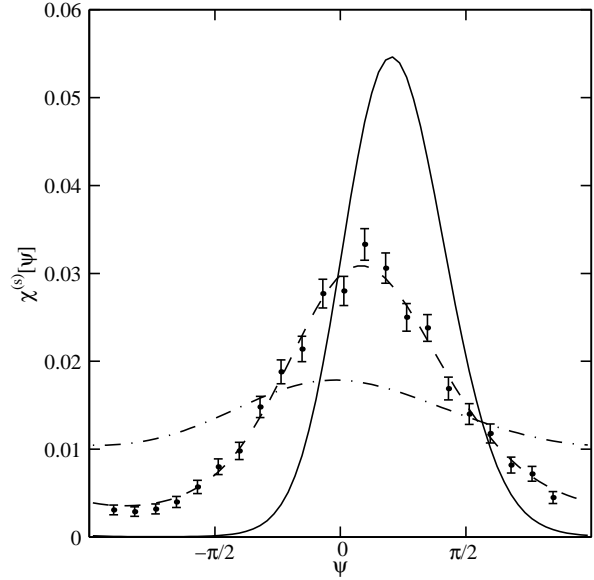}}
  \caption{%
  Stationary phase distributions $\PD{s}$ for stimulus frequency
  $\Omega=0.5\pi$ and noise intensities $D = \eenum{7.8}{-4}$ (solid),
  $D = \eenum{4.8}{-3}$ (dashed), and $D = \eenum{3.0}{-2}$
  (dash-dotted).  Everything else is as in Fig.~\ref{fig:chis_low}.}
\label{fig:chis_high}
\end{figure}

\begin{figure}                   
  \centerline{\epsfbox{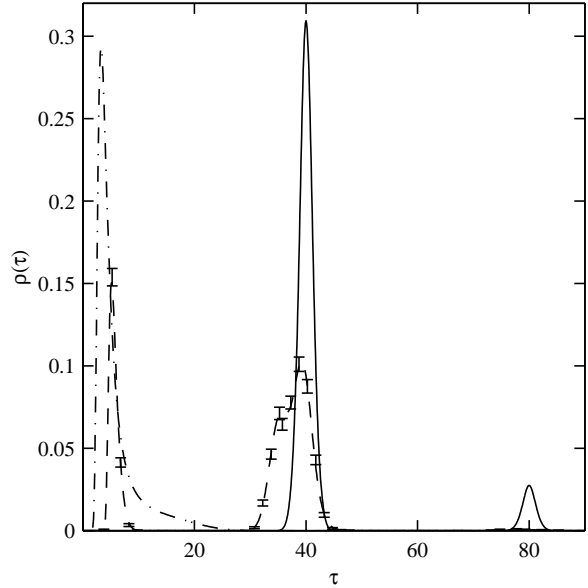}}
  \caption{%
   Stationary ISI distributions for slow stimuli.  Noise intensities are
   $D = \eenum{6.2}{-6}$ (solid),
   $D = \eenum{7.0}{-5}$ (dashed), 
   and $D = \eenum{4.8}{-3}$ (dash-dotted).
   All other parameters are as in Fig.~\ref{fig:chis_low}, error bars
   again indicate simulation results.}
\label{fig:jisi_low}
\end{figure}

\begin{figure}                    
  \centerline{\epsfbox{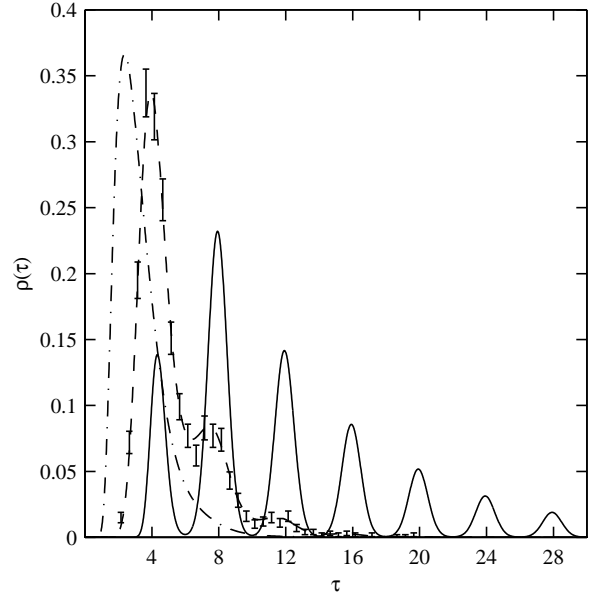}}
  \caption{%
   Stationary ISI distributions for fast stimuli.  Noise intensities are
   $D = \eenum{7.8}{-4}$ (solid),
   $D = \eenum{4.8}{-3}$ (dashed), 
   and $D = \eenum{3.0}{-2}$ (dash-dotted).
   All parameters are as in Fig.~\ref{fig:chis_high}, and error bars
   are from simulations.}
\label{fig:jisi_high}
\end{figure}

Typical ISI distributions $\rho(\tau)$ are given in
Figs.~\ref{fig:jisi_low} and \ref{fig:jisi_high} for the same
parameters as used in Figs.~\ref{fig:chis_low},~\ref{fig:chis_high},
respectively.  For low noise, they contain only periodic peaks,
located precisely at integer multiples of the stimulus period $T$: the
neuron can only fire in a small time window within each period, and
several periods may be skipped in between spikes.  This indicates a
firing pattern that is well phase-locked to the stimulus.  ISI
distributions with comparable structure have been found in neurons of
the auditory system in different
species~\cite{Rose:1967(769),Lavi:1971(467)}.  For high noise, the ISI
distributions reduce to the refractory peak, i.e.\ a largely random
firing pattern.

For intermediate noise, the ISI distributions depend strongly on the
stimulus frequency.  For high frequency (Fig.~\ref{fig:jisi_high}), we
find merely a superposition of periodic and refractory peaks: spikes
preferentially occur at intervals that are multiples of the stimulus
period, but this phase-locking is weak.  This is very different for
slow stimuli (Fig.~\ref{fig:jisi_low}), where the refractory peak is
clearly separated from a wide peak at $\tau=T=40$, the latter exposing
some sub-structure.  This can be understood as follows.  The maximum
of $\rho(\tau)$ at $\tau=T$ corresponds to two spikes fired each at
the optimal phase in two subsequent periods.  In contrast, if a period
that contained a burst of two spikes is followed by another period
containing a burst, then typically the first spike will be slightly
earlier than the optimal phase, the second one a bit later.  Thus, the
interval between the second spike of the first burst and the first
spike of the second burst is shorter than the stimulus period, leading
to the side-peak at $\tau\approx 35$.  The bursts themselves give rise
to the refractory peak.  This again indicates that the spike train is
not a stationary renewal process.

Along with results obtained using the Markov chain approach,
Figs.~\ref{fig:chis_low}--\ref{fig:jisi_high} display phase and ISI
distributions obtained from simulated trains of 20,000 spikes.  The
agreement between Markov model and simulation is excellent.  Source
code for the simulation based on~\cite{Gill:1996(2084)} is available
on request.


\section{Stochastic Resonance}\label{sec:sr}

To assess the performance of the integrate-fire neuron as a signal
processing device, we evaluate the signal-to-noise ratio (\SNR) of the
spike train generated in response to periodic input.  In doing so, one
should keep in mind the purpose of the output spike train.  It has to
convey information to other neurons in the brain \emph{within a
certain time window}, as the brain has to respond quickly to stimuli.
Therefore, the relevant quantity is the signal-to-noise ratio that can
be achieved by measuring the spike train over a finite observation
time $T_o$~\cite{Stem:1996(687)}.

\subsection{Signal-to-noise ratio}\label{ssec:snr}

The one-sided power spectral density of a stationary spike train
$f(t)$ [as defined in Eq.~(\ref{eq:train})] over a time interval $T_o$
is~\cite{Prie:Spec}
\begin{eqnarray}
   S_{T_o}'(\omega) 
   &&=  \frac{1}{\pi T_o}
        \mean{\left| \int_{0}^{T_o} f(t)  e^{i\omega t} dt \right|^2} 
        \nonumber \\
   &&=  \frac{1}{\pi T_o}
        \mean{\sum_{j,k}^{t_j, t_k < T_o} e^{i\omega(t_j-t_k)}} \;.
   \label{eq:psd_T}
\end{eqnarray}
The average is to be taken over the ensemble of all spike trains, that
is, over the set of all conditional ISI distributions and their
$(j-k)$-fold convolutions.  This problem appears intractable.

The situation is greatly simplified if $\omega$ is the stimulus
frequency $\Omega$ or one of its harmonics.  Expressing the spike
times as $t_j = (m_j + \frac{\psi_j}{2\pi})T$, Eq.~(\ref{eq:psd_T})
for $\omega=n\Omega$ simplifies to
\begin{equation}
   S_{T_o}'(n\Omega) =  \frac{1}{\pi T_o}
                        \mean{\sum_{j,k}^{t_j, t_k < T_o} 
                                        e^{i n(\psi_j-\psi_k)}} \;,
   \label{eq:psd_OM}
\end{equation}
where $n, m_j$ are integers, $\psi_j\in[\,0, 2\pi)$, and
$T=2\pi/\Omega$ is the stimulus period.  In the observation period
$T_o$, on average $T_o/\mean{\tau}$ spikes will occur, regardless of
the detailed structure of the spike train.  We therefore fix the upper
limit of the summation at $N_o = \lfloor T_o/\mean{\tau}\rfloor$,
where $\mean{\tau}$ is the average interval length from
Eq.~(\ref{eq:mfpt}) and $\lfloor x \rfloor$ is the largest integer not
exceeding $x$.  This yields as an approximation
\begin{equation}
   S_{T_o}'(n\Omega) \approx S_{T_o}(n\Omega)
        =  \frac{1}{\pi N_o \mean{\tau}}
                        \mean{\sum_{j,k=1}^{N_o} e^{i n(\psi_j-\psi_k)}} \;.
   \label{eq:psd_app}
\end{equation}

The task of computing an expectation with respect to all possible
spike trains is now reduced to that of averaging over all possible
sequences of \emph{spike phases}.  Their distribution and correlations
are completely characterized by the transition matrix $\mat{T}$,
permitting for evaluation of Eq.~(\ref{eq:psd_app}) in closed form.
The actual calculation is straightforward albeit lengthy algebra and
is provided in the appendix.  The final result may be written as
\begin{equation}
  S_{T_o}(n\Omega) 
        = \frac{1}{\pi\mean{\tau}}
               \Bigl[\,1 \,+\, A(n, N_o) \,+\, (N_o-1) B(n)\,\Bigr]
  \label{eq:psd_fin}
\end{equation}
where the functions $A(n, N_0)$ and $B(n)$ are given in the appendix.
Note that $A(n, N_o)$ is bounded as $N_o\rightarrow\infty$.  For a
Poissonian spike train, both $A$ and $B$ are identically zero,
yielding a white power spectrum~\cite{Cox:Stat}.

At first, it might seem surprising that the spectrum contains a term,
$(N_o-1) B(n)$, that scales linearly with the number of spikes in the
train.  This is a consequence of the periodic component of the spike
train introduced by the driving stimulus, leading to a mixed spectrum
consisting of a continuous background and a discrete spectrum of
harmonics~\cite{Prie:Spec}.  For infinite observation time, i.e.\
$N_o\rightarrow\infty$, this gives rise to the terms $\sim
\delta(\omega-n\Omega)$ in the power spectrum.

\begin{figure}                    
  \centerline{\epsfbox{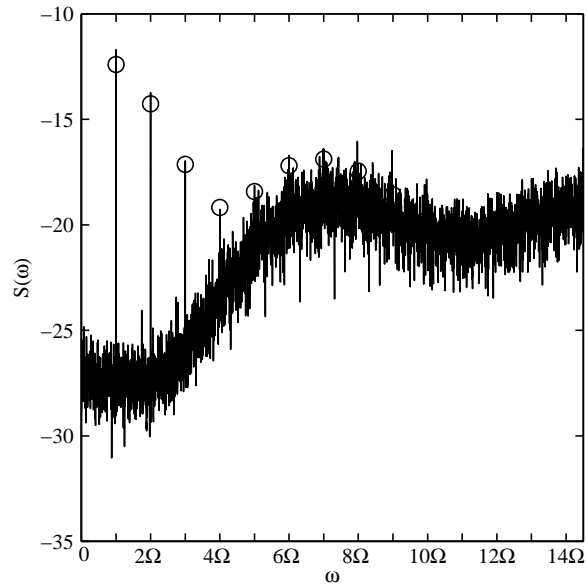}}
  \caption{%
  Power spectral density from an observation time of $T_o=200$ for
  the same stimulus as in Fig.~\ref{fig:chis_low}(b,~e).
  Circles indicate results at stimulus harmonics from the Markov chain
  analysis, while the drawn out line is obtained by FFT from a simulated
  train of 20,000 spikes.  Ticks on the abscissa mark multiples of the
  stimulus period $\Omega=0.05\pi$.}
\label{fig:psd}
\end{figure}

A typical power spectrum is shown in Fig.~\ref{fig:psd}, indicating
close agreement of Eq.~(\ref{eq:psd_fin}) with results obtained by
numerical Fourier transformation of simulated spike trains.  The
approximation made in fixing the summation limit in
Eq.~(\ref{eq:psd_app}) is therefore well justified.  The dip in the
noise background of the spectrum at low frequencies is a consequence
of the refractory period of the neuron, while the weak hump at
$\omega\approx 1$ indicates the presence of
bursts~\cite{Fran:1995(1074)}.  Spectra consisting only of this
background have been found in neurons of higher cortical areas of
monkeys in the absence of periodic input~\cite{Bair:1994(2870)}.

Since the power spectral density can only be evaluated in closed form
at multiples of the stimulus frequency, we approximate the noise
background as Poissonian white noise $S_P = (\pi\mean{\tau})^{-1}$ of
a spike train of equal intensity~\cite{Stem:1996(687)}.  The
signal-to-noise ratio obtainable from the spike train within the
observation time $T_o$ is therefore given by
\begin{equation}
  \SNR_{T_o} = \frac{S_{T_o}(\Omega)}{S_P} 
       = 1 + A(1, \left\lfloor\frac{T_o}{\mean{\tau}}\right\rfloor) 
           + (\left\lfloor\frac{T_o}{\mean{\tau}}\right\rfloor-1) B(1) \;.
  \label{eq:snr_def}
\end{equation}

\begin{figure}                    
  \centerline{\epsfbox{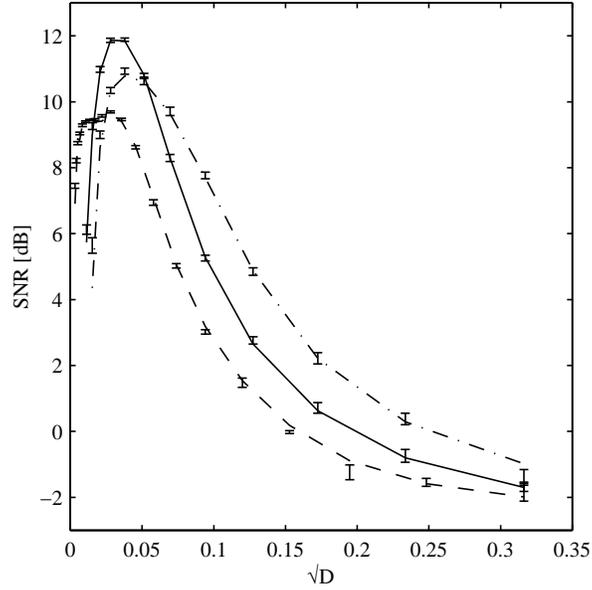}}
  \caption{%
  Signal-to-noise ratio vs.\ noise amplitude for three different
  stimulus frequencies: $\Omega = 0.1\pi$ (dashed), $\Omega =
  0.33\pi$ (solid) and $\Omega = 0.5\pi$ (dash-dotted).  Other parameters
  are $\mu = 0.95$ and $q = 0.05$.  Error bars show
  s.e.m.\ from simulated trains of 20,000 spikes.}
\label{fig:snr_D}
\end{figure}

The signal-to-noise ratio for three different stimulus frequencies is
shown in Fig.~\ref{fig:snr_D} vs.\ the noise amplitude, again in
excellent agreement with simulation results.  Stochastic resonance
(SR) is clearly present at all frequencies, as the \SNR{} attains its
maximum for an intermediate noise level.  The striking new feature is
that the overall maximum in the \SNR{} is reached at an intermediate
frequency $\Omega_r\approx\pi/3$, which we thus call the
\emph{resonance frequency}.  The same qualitative dependence of the
\SNR{} on noise amplitude and stimulus frequencies is observed over a
wide range of stimulus parameters, including weakly supra-threshold
cases ($0.4 \lesssim \mu < 1$, $0.4 \lesssim q / (1-\mu) \lesssim 1.2$;
data not shown).

Note that the stochastic resonance reported in an earlier
paper~\cite{Ples:1997(228)} is an artifact of the renewal ansatz
employed in that work.  There, the stimulus phase is reset to an
arbitrarily chosen value $\phi_0$ after each spike, and the
signal-to-noise ratio is computed for an infinite observation time.
The \SNR{} is maximized for that noise level at which the periodic
peaks of the ISI distribution $\rho(\tau|\phi_0)$ are centered about
the multiples of the stimulus period $T$.  But if, for low noise, one
uses for each noise level $D$ a different $\phi_0(D)$, namely the mode
of the stationary phase distribution as discussed in
Sec.~\ref{ssec:chain}, the periodic peaks are at multiples of $T$ for
all noise intensities, whence the \SNR{} does not drop off for
$D\rightarrow 0$ and no resonance occurs (data not shown).  This
observation underlines the importance of the Markov approach.

\subsection{Time-scale matching}\label{ssec:timescales}

\begin{figure}              
  \centerline{\epsfbox{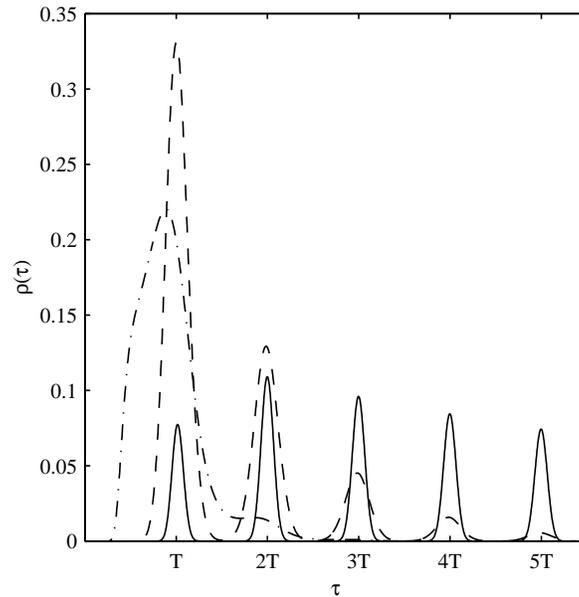}}
  \caption{%
  Inter-spike-interval distributions for the resonance frequency
  $\Omega_r$ and three noise intensities $D=\eenum{1.3}{-4}$ (solid),
  $D=\eenum{7.8}{-4}$ (dashed) and $D=\eenum{4.8}{-3}$ (dash-dotted); 
  other parameters $\mu = 0.95$, $q=0.05$.}
\label{fig:timescales}
\end{figure}

In contrast to stochastic resonance in dynamical systems, SR with
respect to the noise amplitude is not induced by the matching of
time-scales in threshold systems, but results from stochastic
linearization of the response function of the
neuron~\cite{Stem:1996(687),Gamm:1995(4691)}.  In contrast, the
additional resonance along the frequency axis arises in the
integrate-fire neuron as a consequence of matching the stimulus period
to the intrinsic time scale of the neuron in an appropriate manner.
This is demonstrated in Fig.~\ref{fig:timescales}.  For a stimulus at
the resonance frequency $\Omega_r$, the peak at $\tau=T$ in the
stationary ISI distribution can ``grow'' in place as noise is
increased, without being disturbed by the refractory peak.  Indeed,
the latter arises at the location of the first periodic peak and
shifts away from $\tau=T$ only for very large noise.  In this way, the
firing rate of the neuron can be increased without loosing the
phase-locking to the stimulus.  Compare this to the cases of lower
(Fig.~\ref{fig:jisi_low}) and higher (Fig.~\ref{fig:jisi_high})
frequencies: in both cases, high firing rates can only be achieved by
raising the noise amplitude to a point where the refractory peak has
either replaced ($\Omega<\Omega_r$) or smeared out ($\Omega>\Omega_r$)
the periodic peaks, resulting in a firing pattern poorly phase-locked
to the stimulus.

\begin{figure}              
  \centerline{\epsfbox{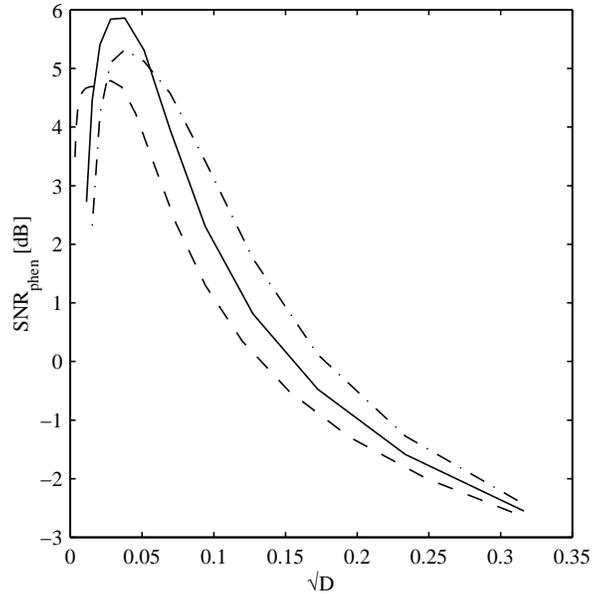}}
  \caption{%
  Signal-to-noise ratio from the phenomenological model of
  Eq.~(\ref{eq:snr_model}).  All parameters are as in
  Fig.~\ref{fig:snr_D}, with stimulus frequencies 
  $\Omega = 0.1\pi$ (dashed), $\Omega = 0.33\pi$ (solid) 
  and $\Omega = 0.5\pi$ (dash-dotted).}
\label{fig:snr_model}
\end{figure}

This competition of precision and intensity is demonstrated by a
phenomenological ansatz for the \SNR{}.  A measure of phase-locking
between stimulus and response is the vector strength $C_s =
\left|\mean{e^{i\psi}}\right|$, where $\psi$ are the spike
phases~\cite{Gold:1969(613)}.  $C_s=1$ indicates perfect and $C_s=0$
no locking.  If the neuron attempts to measure the degree of
phase-locking from a train of $N = T_o/\mean{\tau}$ spikes, the
quality of measurement will be $\sim\sqrt{N}$.  Thus, we expect that
the signal-to-noise ratio will roughly given by
\begin{equation}
  \label{eq:snr_model}
  \SNR_{\text{phen}} \approx {C_s} \sqrt{N} 
          = {C_s}\sqrt{\frac{T_o}{\mean{\tau}}}
  \;.
\end{equation}
Figure~\ref{fig:snr_model} demonstrates that this simple model
describes the behavior of the \SNR{} well.  In particular, the
two-fold stochastic resonance is reproduced.

In short, to elicit a strong output signal from the model neuron, a
sufficient input noise level is required. But this comes at a cost, as
the quality of the output, i.e.\ the precision of the phase locking,
deteriorates as noise is added.  The maximum \SNR{} represents the
optimal compromise between signal strength and quality.


\section{Discussion}\label{sec:disc}

In this paper, we have shown that the periodically driven
integrate-fire neuron can be analyzed in the framework of a Markov
process.  This avoids the unrealistic assumption of a stimulus reset
after each spike, the most serious shortcoming of previous
work~\cite{Buls:1996(3958),Ples:1997(228)}, and this answers
question~(1) raised by Gammaitoni et al.\ in Sec.~V.4.C of their
review~\cite{Gamm:1998(223)}.  Their second questions concerns the
fact that the neural membrane is a rectifier: even a strong negative
input current will not lower the membrane potential more than a few
millivolts below the reset potential $v_0$.  This would indeed be a
problem if the DC offset $\mu$ of the input were much smaller than the
amplitude $q$ of the AC stimulus.  Preliminary evidence suggests that
the best fit of inter-spike-interval distributions generated by the
model with experimental data from the cat's auditory
system~\cite{Lavi:1971(467)} is obtained for sub-threshold stimuli
with $\mu\gg q$.  In this regime, the membrane potential is quickly
raised to $v_0+\mu$ and then oscillates around this level, unaffected
by rectification.  Finally, Gammaitoni and co-authors question the
validity of the approximations used to compute the ISI distributions
in~\cite{Buls:1996(3958)}. This matter is avoided here by numerically
computing these distributions.  A study of the validity of approximate
closed-form ISI distributions will be given
elsewhere~\cite{Ples:1998(NC)}.

The Markov formalism presented in this paper is applicable to any
periodically driven stochastic process with a reset.  The only
required ingredients are the conditional first-passage-time
distributions $\rho(\tau|\phi)$ and the iteration
equations~(\ref{eq:iterate}).  The generalization to more complex
stimuli, e.g.\ including amplitude modulation, is straightforward.

With the Markov machinery at hand, we have demonstrated that the
signal-to-noise ratio of the output of the neuron is maximized at an
optimal noise amplitude for fixed frequency and at a resonance
frequency for fixed noise intensity.  Stochastic resonance with
respect to the stimulus frequency, termed \emph{bona fide} stochastic
resonance, has been described in bistable systems
before~\cite{Gamm:1995(1052),Berd:1996(L447)}.  Therefore, our
findings for a non-dynamical threshold neuron extend the universality
of stochastic resonance to the case of \emph{bona fide} SR.  Recent
criticism~\cite{Choi:1998(6335)} of the original definition of
\emph{bona fide} SR, based on residence time distributions, does not
apply to our study.

Neurons in the auditory system can phase lock to acoustic stimuli with
high acuity and utilize this for the precise localization of sound
sources~\cite{Gers:1996(76)}.  Our results show that strong signals
that are well phase locked to a stimulus may be achieved in spite of
the noise ubiquitous in the neural system.  Stochastic resonance might
therefore be one of the underlying mechanisms of stereo hearing.
First qualitative comparisons indicate good agreement between response
properties of the integrate-fire neuron and of auditory neurons.  An
intriguing question in this respect is the relevance of the \emph{bona
fide} SR to the neural system.  It may serve to tune neurons as
bandpass filters of a special kind: only stimuli in a certain
frequency window will be transmitted with high intensity \emph{and}
precise phase locking.  A detailed study will be the topic of a future
publication.


\acknowledgements 
This work was supported by Deutsche Forschungsgemeinschaft through
SFB~185 ``Nichtlineare Dynamik''.  HEP gratefully acknowledges the
hospitality of the Laboratory for Neural Modeling, Frontier Research
Program, RIKEN, Wako-shi, Saitama, Japan, where this work started.
 

\appendix

\section{Computing the power spectral density}\label{app:psd}

To prove Eq.~(\ref{eq:psd_fin}), i.e.\ 
\begin{eqnarray*}
 S_{T_o}(n\Omega)
        &&=  \frac{1}{\pi M\mean{\tau}}
              \mean{\sum_{j,k=1}^M e^{i n(\psi_j-\psi_k)}} \\
        &&=  \frac{1}{\pi \mean{\tau}}
           \Bigl[\,1 + A(n, M) + (M-1) B(n)\,\Bigr] \;,
\end{eqnarray*}
we split the double sum into the diagonal and off-diagonal terms
\begin{equation}
    S_{T_o}(n\Omega) = \frac{1}{\pi\mean{\tau}}
        \left[1 + h_M(n\Omega) + h^{*}_M(n\Omega)\right] \;,
  \label{eq:S_by_h}
\end{equation}
\begin{equation}
 h_M(n\Omega) = \frac{1}{M} \sum_{k=1}^M\sum_{j=1}^{M-k}
              \mean{ e^{i n (\psi_{k+j} - \psi_k)}} \;,
  \label{eq:def_h}
\end{equation}  
the asterisk denoting complex conjugation and $M=\lfloor
T_o/\tau\rfloor$.

Since we are considering a stationary Markov process, all $\psi_k$ are
identically distributed according to $\PD{s}$, while correlations
between $\psi_k$ and $\psi_{k+j}$ are given by the $j^{\mathrm{th}}$
power of the transition matrix $\mat{T}$ yielding
\begin{equation}
   \mean{e^{i n(\psi_{k+j} - \psi_k)}} 
     =  \vec{\hat{a}}(n)^{\mathrm{tr}} \cdot \mat{T}^j  
          \cdot \vec{\hat{b}}(n)
   \label{eq:mean_einp}
\end{equation}
with vectors 
\[
  \vec{\hat{a}}^{\mathrm{tr}}(n) =
    \big(1,  e^{i n \Dpsi},  e^{2 i n \Dpsi}, \ldots,
          e^{(L-1) i n \Dpsi}\big)
\]
\[
  \vec{\hat{b}}^{\mathrm{tr}}(n)=
    \big(\PD{s}(0),  
         \ldots,  e^{- (L-1) i n \Dpsi}\PD{s}((L-1)\Dpsi)\big)
\]

Upon inserting Eq.~(\ref{eq:mean_einp}) into Eq.~(\ref{eq:def_h}), we
observe that the expression for $h_M$ depends only on $j$ but not on
$k$ so that we may perform the outer summation to obtain
\[
   h_M(n\Omega) = \vec{\hat{a}}^{\mathrm{tr}}(n)
      \bigg[\,\frac{1}{M}\sum_{j=1}^{M-1} (M-j) \mat{T}^j \,\bigg] 
      \vec{\hat{b}}(n) \;.
\]
Diagonalizing $\mat{T}$ leads to
\begin{equation}
    h_M(n\Omega) = 
        \vec{a}(n)^{\text{tr}}\cdot \mat{S}^{(M)}\cdot \vec{b}(n) \;.
  \label{eq:h_by_Sm}
\end{equation}
Here, the diagonal matrix $\mat{S}^{(M)}$ is given by
\begin{equation}
  S_{mm}^{(M)} = 
    \left\{\begin{array}{l@{\qquad}l}
        \displaystyle \frac{M-1}{2} &  \text{for } m = 1 \;,\\
        \displaystyle \frac{\lambda_m}{1-\lambda_m} 
         + \frac{1}{M} \frac{\lambda_m (\lambda_m^M-1)}{(\lambda_m^M-1)^2}
                 &  \text{for}\; m > 1 \;,
           \end{array}\right.
 \label{eq:Sm_def}
\end{equation}
with
\[
   \mat{T} = \mat{C} \cdot  \mat{L} \cdot \mat{C}^{-1} 
   \;, \quad
   \mat{L} = \mathrm{diag}(1 > |\lambda_2| \ge \ldots \ge |\lambda_L|)
   \;,
\]
\[
    \vec{a}(n) = \mat{C}^{\text{tr}} \vec{\hat{a}}(n)
    \;, \quad
    \vec{b}(n) = \mat{C}^{-1} \vec{\hat{b}}(n) \;.
\]
Inserting Eq.~(\ref{eq:h_by_Sm}) into Eq.~(\ref{eq:S_by_h}), we have
\begin{equation}
   S_{T_o}(n\Omega) 
   = \frac{1}{\pi\mean{\tau}}
   \left[1 
     + 2\, \Re\left(\vec{a}^{\text{tr}}(n) \mat{S}^{(M)} \vec{b}(n)\right)
   \right] \;.
\label{eq:S_by_Sm}
\end{equation}

Finally, we split the matrix $\mat{S}^{(M)}$ into the parts pertaining
to the discrete and the continuous parts of the spectrum and define
the functions $A$ and $B$
\[
   \mat{S}^{(M)} = \text{diag}({\textstyle \frac{M-1}{2}}, 0, \ldots, 0)
               + \text{diag}(0, S^{(M)}_{22}, \ldots, S^{(M)}_{LL})  \;,
\]
\[
    A(n, M) = 2\, \Re\left[ \vec{a}^{\text{tr}}(n) \,
                  \text{diag}(0, S^{(M)}_{22}, \ldots, S^{(M)}_{LL}) \,
                  \vec{b}(n) \right] \;,
\]
\[
    B(n) = \Re\left[ a_1(n) b_1(n) \right] \;.
\]
Rewriting Eq.~(\ref{eq:S_by_Sm}) accordingly, we arrive at the desired
expression for the power spectral density
\[
S_{T_o}(n\Omega)
        =  \frac{1}{\pi \mean{\tau}}
           \Bigl[\,1 + A(n, M) + (M-1) B(n)\,\Bigr] \;.
\]

To see that $A$ is bounded as $M\rightarrow\infty$, note that $A$
depends on $M$ only through the diagonal entries of $\mat{S}^{(M)}$
with $m>1$ and $|\lambda_m| < 1$.  For these we have
\[
   \lim_{M\rightarrow\infty} |S^{(M)}_{mm}| = 
       \left|\frac{\lambda_m}{1-\lambda_m}\right| < \infty 
    \;,\quad m > 1 \;.
\]




\end{document}